\documentstyle[twoside,fleqn,espcrc2_pre,espcrc2,epsf]{article}

\newcommand{\x}{\mbox{$\underline{x}$}}

\newcommand{\rr}{\underline{r}}

\newcommand{\rA}{\mbox{$\underline{r}_{A}$}}

\begin{document}

\title{Exotic meson spectroscopy from the clover action at $\beta$ = 5.85 and
6.15\thanks{
Presented by C.\ McNeile.}
}

\author{C.\  Bernard,\address{Department of Physics, 
Washington University, St.\ Louis, MO 63130, USA}
        T.\ DeGrand,\address{Department of Physics, University of
Colorado, Boulder, CO 80309, USA}
        C.\ DeTar,\address{Physics Department, University of Utah, 
        Salt Lake City, UT   84112, USA}
        Steven Gottlieb,\address{Department of Physics,
Indiana University, Bloomington, IN 47405, USA}
        U.\ M.\ Heller,\address{SCRI, The Florida State University,
Tallahassee, FL 32306-4130, USA}
        J.\ Hetrick,\address{Department of physics, University of the
Pacific, Stockton, CA 95211-0197, USA}
        C.\ McNeile${}^c$,
        R.\ Sugar,\address{Department of Physics,
University of California, Santa Barbara, CA 93106, USA}
and
        D.\ Toussaint\address{Department of Physics,
University of Arizona, Tucson, AZ 85721, USA}
}

\begin{abstract}
We repeat our original simulations of the hybrid meson spectrum
using the clover action, as a check on lattice artifacts. Our results
for the $1^{-+}$ masses do not substantially change. We present preliminary 
results for the wave function of the $1^{-+}$ state in Coulomb gauge.
\end{abstract}
 
\maketitle

\section{INTRODUCTION}
There are currently a number of experimental candidates for light
$1^{-+}$ hybrid mesons~\cite{Close:1997ip}. The review of the
experimental results by Page~\cite{Page:1997xs}, suggests that a
$1^{-+}$ state may exist with a mass around $1.6$ GeV.
The predictions of
lattice QCD, by both MILC~\cite{Bernard:1997ib} and
UKQCD~\cite{Lacock:1997ny} predict the lightest $1^{-+}$
state to be around 2 GeV (with large errors). The inclusion of
dynamical fermions~\cite{Lacock:1998} has not 
produced results substantially different
from those of quenched calculations.
To start to check for systematic errors, 
we have repeated our original simulations~\cite{Bernard:1997ib} of the hybrid
meson spectrum that used Wilson fermions, with improved clover quarks.

We have calculated the hybrid meson spectrum  at two different $\beta$
values: $5.85$, with a lattice volume of $20^{3} \times 48$, and
$6.15$ with a lattice volume of $32^{3} \times 64$.
At $\beta=5.85$ we used a clover
coefficient obtained from tadpole improved perturbation theory, using
the plaquette value of $u_0$.  At $\beta=6.15$
we used the non-perturbative value of $c_{sw}$ calculated by
the ALPHA collaboration~\cite{Luscher:1997ug}.
\section{LIGHT QUARK  SPECTROSCOPY}
In Table~\ref{tb:clightresults}, we report our preliminary results for
the mass of the $1^{-+}$ hybrid  in the chiral limit.
At this stage in our analysis, we
have not attempted to estimate the systematic errors in our
results from the clover action.
To set the scale, we used the value of
$r_0$ from the interpolating formulae published
in~\cite{Guagnelli:1998ud} (this changes the number for the Wilson
data slightly from our previously 
quoted number~\cite{Bernard:1997ib}). 
\begin{table}
\begin{tabular}{|c|c|} \hline
Action  & $\kappa_{critical}$     \\  \hline
Wilson  & $ 1980(100)_{stat} \pm sys $  \\
clover  & $ 2110(100)_{stat} \pm sys $  \\
\hline
\end{tabular}
\caption{Mass results for the \protect{$1^{-+}$} state (in MeV) at kappa critical, for the Wilson
and clover actions, at \protect{$\beta = 6.15$}.}
\label{tb:clightresults}
\end{table}
Our preliminary results for the light hybrids
mesons  show that the mass splitting between the $1^{-+}$
and $0^{+-}$ state is considerably reduced, relative to our
previous results~\cite{Bernard:1997ib} 
from the simulations using the Wilson action.
\section{CHARMONIUM SPECTROSCOPY}
We used a more ``traditional''~\cite{Davies:1997hv} 
approach to analysing our hybrid data at the charm mass, than we
originally used in~\cite{Bernard:1997ib}.
In Table~\ref{tb:charmresults}, we present our
results for the mass splittings between the hybrids and the spin
averaged S-wave mass ($M_S = (M_{\eta_c} + 3 M_{J/\psi})/4$), at our
kappa value that corresponds to the charm mass.  Mass splittings
should be less sensitive to lattice
artifacts~\cite{El-Khadra:1992ir,El-Khadra:1997mp}, than the absolute
masses we originally quoted in~\cite{Bernard:1997ib}. 
Also we used the $P-S$
mass splitting to obtain the lattice spacing.
We used the bootstrap method with 100 bootstrap samples to estimate
the errors.

In Table~\ref{tb:charmresults} we see that the clover action
gets the $M_{J/\psi} - M_{\eta_c} $ mass splitting closer to the 
experimental value of $117$ MeV, than the Wilson action. Using potential
model ideas El-Khadra~\cite{El-Khadra:1992ir}
 estimates the value of this splitting in the
quenched approximation to be $\sim 70 $ MeV.

The importance of the heavy-heavy hybrid mesons, and other
approaches to studying them, is  discussed
by Kuti~\cite{Kuti:1998}.
\begin{table}
\begin{tabular}{|c|c|c|} \hline
Quantity & Wilson & clover \\  \hline
$a^{-1}_{P-S} $  & $2650^{+110}_{-90}$ & $2900_{-90}^{+90}$ \\ 
$M_{J/\psi} - M_{\eta_c} $  & $27_{-1}^{+1} \pm sys $ & $71_{-2}^{+2} \pm sys$ \\ 
$M_{1^{-+}} - M_S$  & $1340_{-150}^{+60} \pm sys$ & $1220_{-190}^{+110} \pm sys
$ \\
$M_{0^{+-}} - M_S$  & $1490_{-100}^{+110} \pm sys $ & $1490_{-110}^{+80} \pm sys$ \\
\hline
\end{tabular}
\caption{Mass splitting results (in MeV) for charmonium for the Wilson
and clover actions
at $\beta = 6.15$.}
\label{tb:charmresults}
\vskip -8mm
\end{table}
\section{WAVE FUNCTIONS}
To study the internal distribution of quarks and glue
inside a hybrid meson, we measured
\begin{equation}
    \sum_{\x}  \overline{q}(\x,t)  F(\x + \rA + \rr,t) \Gamma q(\x+\rr,t)
\label{eq:hybdefn}
\end{equation}
where $F$ is the field strength tensor and we have suppressed the 
contraction of $F$ with the $\Gamma$  gamma matrix.
We fix to Coulomb gauge, measure a correlator with Eq.~\ref{eq:hybdefn}
at the sink with our standard hybrid meson source~\cite{Bernard:1997ib} at time slice $t=0$.
The operator in Eq.~\ref{eq:hybdefn} is similar to the one used
to measure the wave function of the proton~\cite{Hecht:1992uq}, except
that one of the quarks is replaced by the field strength tensor.
Although the wave function operator is difficult to interpret in
terms
of constituent gluons, it makes sense in terms of a Fock space analysis.

\begin{figure}
\vbox{\epsfxsize=2.7in \epsfbox{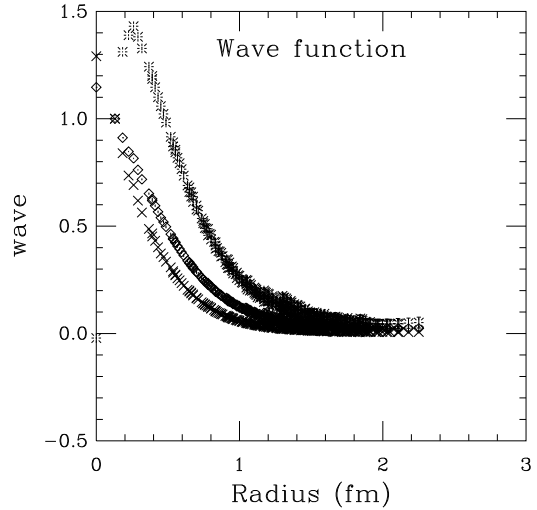}}
\vskip -9mm
\caption{Wave function of the pion (crosses), rho (diamonds) and
\protect{$1^{-+} $} 
state (bursts), at $\beta = 5.85$.}
\label{fig:wave}
\vskip -8mm
\end{figure}

For our preliminary results we use $\rA = 0$. However, it will be
interesting to  measure the wave function with non-zero $\rA$ in
Eq.~\ref{eq:hybdefn}.  
One approach to studying the heavy hybrids is to solve the
Schr\"{o}dinger  equation for the quarks with an excited 
potential, measured in a lattice simulation~\cite{Kuti:1998}. The resulting
wave functions do not have any dependence on $\rA$.
A study of the dependence of the operator in
Eq.~\ref{eq:hybdefn}, on $\rA$, may provide insight into the validity
of the excited potential approach to studying hybrid mesons.

In Fig.~\ref{fig:wave} we plot the wave functions of the pion, rho and $1^{-+}$
states in Coulomb gauge, at $\beta = 5.85$ and  $\kappa = 0.135$
(corresponding to a vector to pseudoscalar mass ratio of $0.85$), with
a sample size of 40. The scale on the x axis is set using the chirally
extrapolated rho mass.

The wave function in Fig.~\ref{fig:wave}  looks qualitatively similar to those 
obtained~\cite{Juge:1997nc} by solving the Schr\"{o}dinger equation 
with an excited lattice potential
(although we are working with lighter quarks).

\section{4-QUARK HYBRID MIXING}
An important issue in the spectroscopy of hybrid mesons is to
understand the mixing between a hybrid meson and a 4-quark state with 
the same quantum numbers. We studied this issue by using the valence
approximation.
The (naive) valence approximation~\cite{Liu:1994cv} removes
\begin{equation}
-\kappa ( 1 - \gamma_4 ) U_4(x) \delta_{x,y-\hat{t}}
\label{eq:valence}
\end{equation}
from the quark action, so that the quarks travel forwards 
in time only (note that NRQCD~\cite{Thacker91a} 
quarks only propagate forwards in time
as well).
In the quenched approximation, it is this term that causes 
mixing between hybrid and 4-quark states 
via hairpin diagrams (see Fig. 1 in~\cite{Bernard:1997ib}).
Liu and collaborators~\cite{Liu:1998um} have developed a valence approximation
with an improved non-relativistic limit over the prescription in
Eq.~\ref{eq:valence} suitable for the investigation of the 
relationship between the quark model and QCD. However, for our
purposes, the removal of Eq.~\ref{eq:valence} from the 
clover action is adequate to study the effect of the hairpin
graph on the hybrid spectrum.

\begin{figure}
\vbox{\epsfxsize=2.7in \epsfbox{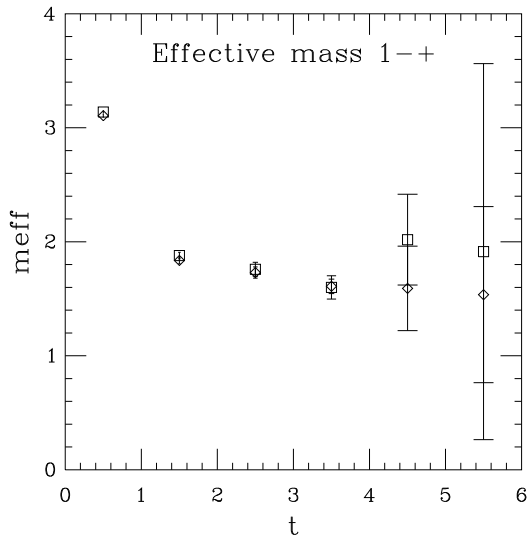}}
\vskip -9mm
\caption{Effective mass plot for the \protect{$1^{-+}$} hybrid in the 
valence (squares) and quenched (diamonds) approximations, at
\protect{$\kappa = 0.135$}.}
\label{fig:valence}
\vskip -8mm
\end{figure}

In this preliminary study, Fig.~\ref{fig:valence} shows that the
effective mass plots of the $1^{-+}$ state, in the valence and
quenched approximations, look very similar. This suggests that at the
parameters
at which we are working,
mixing via hairpin diagrams between the
$1^{-+}$ hybrid operator and four quark states is a small effect.

%
%

This work is supported by the DOE and the NSF.  The computations were
carried out at CCS (ORNL), CHPC (Utah), and NPACI (SDSC).



\end{document}